\documentclass[twocolumn,showpacs,preprintnumbers,amsmath,amssymb,nofootinbib]{revtex4}

\usepackage{graphicx}
\usepackage{dcolumn}
\usepackage{bm}

\newcommand{\MeV}{~\mathrm{MeV}}
\newcommand{\GeV}{~\mathrm{GeV}}
\newcommand{\TeV}{~\mathrm{TeV}}
\newcommand{\zp}{Z^{\prime}}
\newcommand{\uonep}{U(1)^\prime}
\newcommand{\mzp}{M_{\tilde{Z}^\prime}}
\newcommand{\zpino}{\tilde{Z}^\prime}
\newcommand{\gz}{g_{z^\prime}}

\newcommand{\svev}{\langle S \rangle}
\newcommand{\mzpr}{M_{\zp}}
\newcommand{\lams}{\Lambda_S}

\begin{document}


\title{Aspects of Z$^\prime$-mediated Supersymmetry Breaking} 

\author{Paul Langacker$^*$, Gil Paz$^*$, Lian-Tao Wang$^\dagger$, Itay
  Yavin$^\dagger$} 
\affiliation{%
$^*$ School of Natural Sciences, Institute for Advanced Study,
  Einstein Drive Princeton, NJ 08540 \\ 
$^\dagger$ Physics Department, Princeton University,  \\ Princeton NJ 08544 
}%

\date{\today}

\begin{abstract}
In a recent paper, we proposed the possibility that
supersymmetry breaking is  
communicated dominantly via a $\uonep$ vector multiplet. We also required
that the $\uonep$ plays a crucial role in solving the $\mu$
problem. We discuss here in detail both the construction and the
phenomenology of one class of such models.  The low energy
spectrum generically   
contains heavy sfermions, Higgsinos and exotics $\sim 10-100\TeV$;
an intermediate $M_{\zp}$ $\sim 3-30\TeV$; light gauginos $\sim 
100-1000\GeV$, of which the lightest can be wino-like; a light
Higgs with a mass of  $\sim140\GeV$; and a singlino which can be very light.  We present a set of possible
consistent charge choices. Several benchmark models are used to
demonstrate characteristic phenomenological features. 
Special attention is devoted to
interesting LHC signatures such as gluino decay and the decay patterns
of the electroweak-inos. 
Implications for  neutrino 
masses, exotic decays, $R$-parity, gauge unification, and the gravitino mass are briefly discussed.
\end{abstract}

\pacs{12.60.Jv, 12.60.Cn, 12.60.Fr, 14.80Ly}
\maketitle

\section{\label{sec:intro}Introduction}

Many supersymmetry breaking mediation mechanisms, such as gravity 
mediation~\cite{Chamseddine:1982jx,Barbieri:1982eh,Nilles:1982dy,Cremmer:1982vy,Ohta:1982wn,Hall:1983iz,Soni:1983rm},   
anomaly  mediation  ~\cite{Randall:1998uk,Giudice:1998xp}, gauge
mediation ~\cite{Dine:1981za}-\cite{Giudice:1998bp}, and gaugino mediation 
\cite{Chacko:1999mi,Kaplan:1999ac}, have been proposed (for a review,
see~\cite{Chung:2003fi}).  In a recent paper 
\cite{zpmed_short}, we proposed that supersymmetry breaking could
instead be communicated naturally by some exotic gauge interactions. A
typical example of such a mediator  is an extra $\uonep$\footnote{Scenarios involving
an extra $\uonep$ in supersymmetry mediation have been considered previously~\cite{Dobrescu:1997qc}-\cite{Everett:2000hb}.
Here we assume that the $\zp$-mediation is the dominant source for both scalar and gaugino masses.}. The
existence of low energy supersymmetry would give indirect evidence
that TeV scale new physics could be directly embedded into some high
scale fundamental theory, such as string theory. Concrete
semi-realistic superstring constructions frequently lead to additional
non-anomalous $\uonep$ factors in the low-energy theory (see, e.g.,
\cite{Faraggi:1989ka}-\cite{Binetruy:2005ez}), and in some cases both
the ordinary sector and hidden sector particles carry $\uonep$
charges, allowing a $\uonep$-mediated communication between the two
sectors. More recently \cite{Verlinde:2007qk}, it was realized that
there is a natural way of implementing such a mediation 
mechanism in a large class of D-brane constructions.

Motivated by the $\mu$-problem of the MSSM, we focused on one class of
solutions, which invokes a spontaneously broken PQ symmetry
(see, e.g., ~\cite{Accomando:2006ga}). From the
point of view of top-down  constructions it is common that such a
symmetry is promoted to a $U(1)'$ gauge symmetry~
\cite{Suematsu:1994qm,Cvetic:1997ky}. It is natural to
make this $\uonep$ the mediator of SUSY breaking as well, since in this case
$\mu$ (as well as $\mu B$) will be set by the scale of the other soft
SUSY breaking parameters. Whether or not  the electroweak symmetry
breaking is finely tuned, $\mu$ and $\mu B$ terms generated this way
are of the right-size. We would like to  include this as a feature of
the class of models we consider, though it is not absolutely essential.

In our setup, a supersymmetry breaking $\zp$-ino mass term, $\mzp$, is
generated due 
to $\uonep$ coupling to the hidden sector. The observable sector
fields feel the supersymmetry breaking through their couplings to
$\uonep$, implying interesting features of the sparticle spectrum. The
sfermion masses are of the  order of $m_{\tilde{f}}^2 \sim
\mzp^2/16 \pi^2$. The $SU(3)_C \times SU(2)_L \times U(1)_Y$ gaugino
masses are generated at higher loop order, $M_{1/2} \sim  \mzp /(16
\pi^2)^2$, which is 2-3 orders of magnitudes lighter than the
sfermions.   LEP direct searches suggest electroweak-ino  
masses $> 100$ GeV \cite{lepsusy}. We
therefore expect that  the sfermions are heavy, typically
about $100$ TeV.
In this sense, this scenario could be viewed as a mini-version of
split-supersymmetry~\cite{ArkaniHamed:2004fb,ArkaniHamed:2004yi}. One
important difference is the $\mu$-parameter, which
is set by the scale of $\uonep$ breaking. Although in principle a free
parameter, we find it is naturally at the same order of magnitude as
the sfermions. Similar to split supersymmetry, one fine-tuning is
needed to maintain a low electroweak scale. The scenario does not have
flavor or CP violation problems due to the decoupling of the
sfermions. The flavor violation in the scenario will be further
suppressed if we choose flavor universal $U(1)'$ charges for the
Standard Model matter fields. Due to the same decoupling effect, we
expect that the contribution to the muon anomalous magnetic moment is
negligible in this scenario. 

This paper is organized as follows. We first review the generic
setup and resulting sparticle spectrum. Then, as an example,  we
construct a specific
model with more assumptions about the consistency conditions and the
existence of specific types of exotics. Finally, we comment on the
phenomenology of this class of models, including the spectrum, the gluino lifetime, cold
dark matter,
possible ranges for the gravitino mass, exotic
decays, possibilities for neutrino mass, $R$-parity, and gauge unification.

\section{Generic Features of Z$^\prime$-mediated Supersymmetry
  Breaking}
\label{sec:generic}

\begin{figure}
\begin{center}
\includegraphics[scale=1.0]{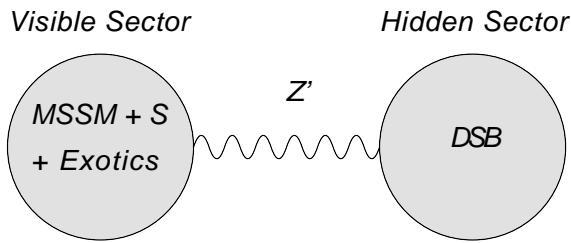}
\end{center}
\caption{Z$^\prime$-mediated supersymmetry breaking.}
\label{fig:two-sectors}
\end{figure}

The schematics of the $\uonep$ mediation model is presented in
Fig. \ref{fig:two-sectors}.
We extend the MSSM in the following ways. First, introduce an extra
$U(1)^{\prime}$ gauge symmetry. Second, promote the $\mu$ parameter
into a dynamical field, $S$, which is charged under the
$U(1)^{\prime}$. Third,  include other exotics with Yukawa couplings
to $S$. The last assumption is included to drive the necessary
radiative symmetry breaking and to cancel anomalies. Such exotics and
couplings generically exist in string theory constructions. The
superpotential  is
\begin{eqnarray}
\label{eqn:superpotential}
W &=& y_u  {H}_u {Q} {u}^c + y_d  {H}_d {Q} {d}^c + y_e  {H}_d {L} {e}^c
\\\nonumber
&+& y_\nu  {H}_u {L} {\nu}^c +\lambda  {S}  {H}_u  {H}_d
\\\nonumber
&+& \sum_{i \in \{ \mbox{exotics} \}} y_i {S} X_i X^{c}_i,
\end{eqnarray}
where $i$ labels the species of exotics.

\subsection{Features of the Spectrum}
We begin by discussing the pattern of the soft supersymmetry breaking parameters, the masses of the
$\zp$-ino and of the MSSM squarks and gauginos, which are the most robust predictions of this scenario.
At the supersymmetry breaking scale, $\Lambda_S$, supersymmetry
breaking in the hidden sector is assumed to generate a supersymmetry
breaking mass for the fermionic component of the $U(1)'$ vector superfield.
 Given details of the hidden sector, its value could
be evaluated via the standard technique of analytical continuation
into superspace \cite{ArkaniHamed:1998kj}. In particular, the gauge
kinetic function of
the field strength superfield $\widehat{Z}^\prime$ at the supersymmetry
breaking scale is  
\begin{eqnarray}
\label{eqn:zpinomass}
\mathcal{L}_{\rm \zpino} &=& \int d^2 \theta \left[ \frac{1}{\gz^2
  (0)} + \beta_{\zp}^{hid} \log
  \left( \frac{\Lambda_S}{M} \right) \right. \nonumber \\ &+& \left.
 \beta^{vis}_{\zp}\log \left(
  \frac{\Lambda_S}{\mzp} \right) \right] \widehat{Z}^{\prime} \widehat{Z}^{\prime},
\end{eqnarray}
where $M$ is the messenger scale, which we have assumed to be around
the supersymmetry breaking scale, $M \sim
\Lambda_S$. $\beta_{\zp}^{hid}$ and $\beta^{vis}_{\zp}$ are
$\beta$-functions induced by $U(1)'$ couplings to hidden and visible
sector fields, respectively.    Using analytical
continuation, we replace $M$
with $M+ \theta^2 F$, where $F$ is the supersymmetry breaking order
parameter.  We obtain the $\zpino$ mass as $ \mzp \sim \gz^2
\beta_{\zp}^{hid} F/M $. We assume that the $\uonep$ gauge symmetry is not broken in the hidden sector.

We assume that  all the chiral superfields in the visible sector are
charged under $\uonep$, so all the corresponding scalars receive  soft
mass terms at 1-loop of order\footnote{
  Eq. (\ref{eqn:scalarmass}) cannot be the full story or we would not
  be able to drive the singlet scalar mass-square negative or keep the Higgs
  light. However, this contribution does serve to set the overall
  scale. To generate a much lighter mass scale requires fine-tuning.},
\begin{equation}
\label{eqn:scalarmass}
m^2_{\tilde{f}_i} \sim \frac{\gz^2 Q_{f_i}^2}{16 \pi^2} \mzp^2
\log\left(\frac{\Lambda_S}{\mzp} \right),
\end{equation}
where $\gz$ is the $\uonep$ gauge coupling and $ Q_{f_i}$ is the
$\uonep$ charge of $f_i$, which we take to be of order 1. (The exact
expressions can be determined from the renormalization group
equations (RGEs) given in Appendix  \ref{app:RGEs}.)

The $SU(3)_C \times SU(2)_L \times U(1)_Y$ gaugino masses, however,
can only be generated at 2-loop level
since they do not directly couple to the $\uonep$,
\begin{eqnarray}
\label{eqn:gauginomass}
&\quad& \nonumber \\
{M}_a~~ &\sim& \parbox[t]{5cm}{\vspace{-1cm}
  \includegraphics[scale=0.45]{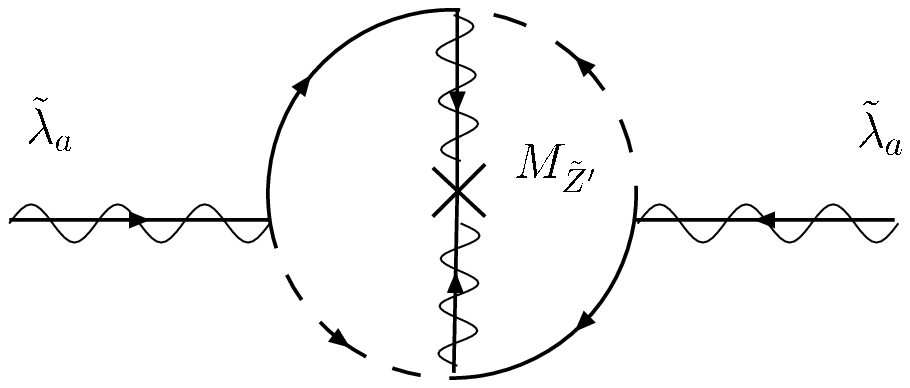} }  \nonumber \\
&\sim& ~~\frac{\gz^2 g_a^2}{(16\pi^2)^2} \mzp
\log\left(\frac{\Lambda_S}{\mzp} \right),
\end{eqnarray}
where $g_a$ is the gauge coupling for the gaugino $\tilde{\lambda}_a$, and the
internal line is the sum over the
chiral supermultiplets charged under the $a^{th}$ gauge group.
(We have suppressed the group and  $\uonep$ charge factors.)
Since these gaugino masses are proportional to $g_a^2$, we expect that the
gluino will typically be significantly heavier than the  others. However, that conclusion and the
ordering of wino and bino masses depends on specific charge
assignments and the exotic matter content.

 From the discussion above, we see that the
gauginos are considerably lighter than the sfermions.
Taking  ${M}_a\gtrsim 100\GeV$, we find
\begin{equation}
 \mzp \log\left(\frac{\Lambda_S}{\mzp} \right) \sim 10^4 \TeV
 \end{equation}
 and
\begin{equation}
{m}_{\tilde{f}_i} \sim \frac{(4\pi)^3}{\gz g_a^2} M_a \sim 100\TeV,
\end{equation}
where we have assumed that $\gz$ is of electroweak strength.
Similarly, $A$ terms associated with the Yukawa interactions in
(\ref{eqn:superpotential})
are generated at one-loop by $\zpino$ exchange, yielding
\begin{equation}
A \sim  \frac{y \gz^2 }{16 \pi^2} \mzp \log\left(\frac{\Lambda_S}{\mzp} \right) \sim y \times  10\TeV,
\end{equation}
where the Yukawa coupling $y$ is absorbed into $A$. Again, the exact expressions, including
the counting factors and dependence on the $\uonep$ charges, can be found from the
expressions in  Appendix  \ref{app:RGEs}.

The gravitino mass $m_{3/2} \sim F/ M_P$ depends strongly on the size
of supersymmetry breaking in the hidden sector. Requiring MSSM gaugino masses
$\gtrsim$ 100 GeV sets $F/M \geq 10^6 /( \gz^2 g_a^2)$ GeV. Assuming
$\sqrt{F}$, $M$ and $\lams$ to be within an order of magnitude,
we could have $\sqrt{F} \sim 10^7 -10^{11}$ GeV. This gives a wide
range of gravitino masses with very different phenomenologies, as will
be discussed in Section.~\ref{sec:gravitino}. This is very different
from gauge mediated supersymmetry breaking, where a typically lower
supersymmetry breaking scale ($\sim 10 - 1000$ TeV) implies
a gravitino much lighter than the other superpartners. Without
specifying a particular model of supersymmetry breaking and gravity
mediation, we will treat it as a free parameter to begin with. The
gravitino will be further constrained by cosmological data such as Big
Bang Nucleosynthesis and by the cold dark matter density $\Omega_{CDM}$,
which we will discuss in Section~\ref{sec:gravitino}. We will assume
that $\sqrt{F}$ is not very different, within a couple of orders of
magnitude, from the supersymmetry breaking scale $\lams$. In the
scenario under consideration, the supersymmetry breaking scale is
constrained logarithmically by the requirement of radiative symmetry
breaking. 
Since the relevant Yukawa contributions that ultimately fix the electroweak scale are
proportional to $ y_i^2 \log (\lams/ \mzp ) /16 \pi^2$,
the  gravitino mass is exponentially sensitive
to the choice of the parameters in the model, as $m_{3/2} \propto
e^{1/y_i^2}$.

\subsection{Kinetic mixing}

Kinetic mixing between the
$\zp$-ino and bino will be generated at one loop level through interactions
with the visible sector matter content. It is generically of the
order $k\sim(\gz g_Y tr \left(Q_i Y_i\right)/16 \pi^2) \log (\Lambda_S /
\mzp)$ (where $Q_i$ are the $\uonep$ charges, $Y_i$ are the
hypercharges and the trace is taken over all charged matter). The
existence of such kinetic mixing implies that we must add a term $
K \widehat{B} \widehat{Z}^{\prime} $ to Eq.~\ref{eqn:zpinomass}, where $K$
is in general  a holomorphic function whose lowest component acquires
a vev $k$. If $K$ involves some hidden sector field, the induced
correction to the light bino mass will be on the order of $k^2 \mzp$,
which is at the same order as the contribution considered in the
previous section. However, we have assumed that only visible sector
fields, which do not participate in supersymmetry breaking, are
charged under  both hypercharge and $\uonep$. Therefore, by construction, such a
contribution is absent at one loop level in our scenario. It will
enter at higher loop order, which is negligible in comparison with the
two-loop contribution we have considered. 

We now discuss the effect of the kinetic mixing. 
This will shift the mass of the bino, but such a shift is
proportional to the square of the bino's mass.

The gaugino kinetic and mass terms are
\begin{eqnarray}
\mathcal{L}&=& -i\left(\widetilde{B\mbox{}}^\dagger\, \tilde{Z'}^{\dagger}\right)
\left(\begin{array}{cc}1&k\\k&1\end{array}\right)\bar\sigma^\mu D_\mu
\left(\begin{array}{c}\widetilde{B}\\\tilde{Z'}\end{array}\right)\\\nonumber
&&+
(\widetilde{B\mbox{}}\,\,
\tilde{Z'})\left(\begin{array}{cc}M_1&0\\0&\mzp\end{array}\right) 
\left(\begin{array}{c}\widetilde{B}\\\tilde{Z'}\end{array}\right),
\end{eqnarray}
where we have ignored non-diagonal terms which are of the same order
as $M_1$ due to their negligible effect on the bino mass.  
After bringing the kinetic term to its canonical form, we find that
the new mass eigenvalues are
\begin{equation}
M_1\to M_1(1-\frac{M_1k^2}{\mzp}),\quad  \mzp\to\mzp(1+k^2),
\end{equation}
where $M_1 \ll \mzp$, $k < 1$, and we keep only the lowest order terms. This
result can be understood in terms of chiral symmetry. In the limit of
 vanishing bino mass, the zero eigenvalue of the gaugino mass
matrix is not changed by the congruence transformation that brings the
kinetic term to its canonical form. As a result, once the bino mass
is ``turned on", the shift must be proportional to it. 

Similar results apply to kinetic mixing between the $Z$ and $Z'$ gauge
bosons \cite{Holdom:1985ag,Babu:1997st}. In particular when the $Z'$
gauge boson becomes massive, the $Z$ gauge boson remains massless. The
only effect this mixing will have is to shift the $Z'$ gauge coupling:
$\gz Q_i\to \gz Q_i-kg_Y Y_i$.

\subsection{Contribution from other mediation mechanisms}

Since the soft parameters
resulting from $\zp$ mediation involve a large hierarchy,
non-dominant contributions   from other mediation mechanisms could also
be important. For example, there
could be other gauge interactions between the hidden and observable
sector. However, as long as they do not contain SM gauge interactions,
we expect the main features of the soft breaking parameters to continue
to hold, as most of our discussion above is independent of the gauge
group (except the kinetic mixing between $U(1)$ factors). The other
obvious candidate is gravity mediation, which yields a contribution to
the gaugino mass of order $F/M_P$.  Since
we have $\mzp \sim (\gz^2 /16 \pi^2) F/M$, we expect $\sqrt{F} \sim
10^7 - 10^{11}$ GeV, without assuming a large hierarchy between
$\Lambda_S$, $\sqrt{F}$ and $M$. Therefore, gravity mediation could
give comparable contributions to the gaugino masses for higher values
of $F$. On the other hand, its contribution to soft scalar
mass-squares $\sim F^2/M_P^2$ is 
expected to be much smaller than the $\zp$-mediated contribution. This
is very different from gauge mediation through the Standard Model
gauge groups, where all the soft terms are of the same order.
Therefore, while in principle the gravity mediation contribution to
the gaugino masses could be comparable to the one from
$\zp$-mediation, we expect the hierarchy between the sfermions and
scalars to be a robust prediction of this scenario.

It is possible that the gravity mediation piece is
sequestered \cite{Sundrum:2004un,Schmaltz:2006qs,Kachru:2007xp}. In
this case, the dominant supergravity contribution will come from
anomaly mediation. Such contribution could be important in our case if
$\Lambda_S \sim 10^{11} - 10^{12}$ GeV.

\subsection{Symmetry breaking and fine-tuning}
\label{sec:symmetry-breaking}

The $\uonep$ gauge symmetry is broken by the vev $\langle S
\rangle$. We assume that this symmetry breaking is
triggered by radiative corrections to the soft mass $m_S^2$,
especially through Yukawa 
couplings to exotics\footnote{An alternative possibility, which we
  have not investigated, 
would be for the $\uonep$ gauge symmetry to also be broken in the
hidden sector. In that case, 
$\mzp$, $\mzpr$, and $\lams$ would all be free parameters.}.

We are looking for parameters which result in solutions such that $\svev \gg v$,
where $v\equiv ( |\langle H_u^0 \rangle|^2+  |\langle H_d^0 \rangle|^2 )^{1/2} \sim 174\GeV$ is the electroweak scale.
It is therefore reasonable to first determine $\svev$ ignoring the
Higgs doublets, and
then to consider the Higgs potential for the doublets regarding
$\svev$ as a fixed parameter.
We have verified that the corrections from the shift in $\svev$ due to
the doublets is small.
The scalar potential for $S$ is
\begin{equation}
\label{eqn:Spotential}
V(S) = m_S^2 |S|^2+ \frac{1}{2} \gz^2 Q_S^2 |S|^4,
\end{equation}
which is minimized for
$\svev^2 ={-m_S^2}/{\gz^2 Q_S^2}$ for $m_S^2 <0$.
The $\uonep$ symmetry breaking is driven by the radiative corrections
to $m_S^2$.
\begin{eqnarray}
16\pi^2 \frac{d m_S^2}{d \log \mu} &=& -8 \gz^2 Q_S^2 \mzp^2  \\\nonumber
&+& 4\lambda^2 (m_S^2 + m_{H_u}^2 + m_{H_d}^2) \\ \nonumber  &+&2 \sum_{
\{ \mbox{exotics} \}} y_i^2(m_S^2
+ {m}_{X_i}^2 + {m}_{X^{c}_i}^2 ).
\end{eqnarray}
The charges and Yukawa couplings have to be chosen so that radiative
symmetry breaking actually occurs. The relative contribution to
$m_S^2$ from the exotics goes as $- (y_i^2/16 \pi^2)\,m^2_{\tilde{f}_i} \log (\Lambda_S /
\mzp) $ \footnote{The Yukawa contribution to
the running actually continues below $\mzp$, but in most of the cases
considered this is a small effect.}. Therefore, successful radiative
breaking of $\uonep$ usually requires that the Yukawa couplings to the
exotics are not small and  that some hierarchy exists between
$\Lambda_S$ and $\mzp$, which depends on the choice of the Yukawa
couplings. We will illustrate such effects in the context of a
specific model, for which we typically find $ \langle S \rangle \sim 100\TeV$.

 Meanwhile,  to generate the electroweak scale we must fine-tune one
 linear combination of the two Higgs doublets to be much lighter than
 its natural scale. The $\zp$-ino mass, $\mzp$, sets the overall scale
 in the visible sector, so the tuning must be between the
 dimensionless couplings in the model, namely $\gz$, $\lambda$,
and the other Yukawas, as well as the ratio $\log(\Lambda_S/{\mzp})$.
While the restriction on the parameter space from $\uonep$ breaking is
 model dependent, the need for fine-tuning to obtain the electroweak symmetry breaking is
 generic.

The full mass matrix for the two Higgs doublets is,

\begin{eqnarray}
\label{eqn:higgsMatrix}
\mathcal{M}_H^2 &=& \left(
\begin{array}{cc}
m_{2}^2 & -
A_H \svev \\
\\
- A_H \svev & m_{1}^2
\end{array}
\right) \nonumber \\
m_{2}^2 &=& m_{H_u}^2 + \gz^2 Q_S Q_2 \svev^2 + \lambda^2 \svev^2
\nonumber \\
m_{1}^2 &=& m_{H_d}^2 + \gz^2 Q_S Q_1 \svev^2 +
\lambda^2 \svev^2,
\end{eqnarray}
where $Q_2\equiv Q_{H_u}$, $Q_1\equiv Q_{H_d}$, and all the couplings
and mass terms are evaluated at $\mzp$. $A_H$ is the radiatively
generated soft trilinear coupling between the Higgs doublets and the
singlet, $\mathcal{L}=- A_H H_u H_d S +\text{h.c.}$. 
Electroweak breaking requires one small eigenvalue $\mathcal{O}
(v^2)$.
Since $A_H$ has
unit mass dimension it is generally 
suppressed with respect to the scalar soft masses by about an order
of magnitude ($\sim 4\pi$). Therefore, we will have to tune one of the
diagonal terms to be small. The up-type Higgs soft mass can usually be
driven negative owing to the large top Yukawa coupling. We can then
tune it against the other contributions to the diagonal up-type
entry. In particular, we will adopt the following scheme for the
tuning. Since $\svev^2$ has a different dependence on $\lams$ than
$m_{H_u}^2$ we will keep all the couplings and mass
scales fixed and allow ourselves to vary only $\lams$. This suffices
to generate a small eigenvalue and obtain the electroweak scale. 

Since the down-type mass term is much larger than all the other
scales, $\tan\beta$ is well approximated by,
\begin{equation}
\tan\beta = \frac{m_{1}^2}{A_H \svev} \sim
10-100. 
\end{equation}

Next, we turn to discuss the part of the mass spectrum which will be
determined by the $\uonep$ symmetry breaking.

The effective $\mu$ term is $\mu = \lambda \svev$. Assuming $\lambda=\mathcal{O}(0.1-1)$, we have
$\mu \sim $  10-100 TeV. Similarly, the fermionic component of the exotic superfields $X_i$ and $X_i^c$ will acquire
supersymmetric masses $y_i \svev \sim $ 10-100 TeV.

The $\zp$ mass is
\begin{equation}
M_{\zp}= \sqrt{2}\gz Q_S \svev=\sqrt{2}|m_S|.
\label{eqn:zpmass}
\end{equation}
The singlino $\widetilde{S}$ receives a mass through mixing with the
$\zp$-ino. The mass matrix is given by,
\begin{equation}
\label{eqn:SZmixing}
\mathcal{M}_{SZ} = \left(
\begin{array}{cc}
0 &  \sqrt{2}\gz Q_S \svev \\
\\
\sqrt{2}\gz Q_S \svev & \mzp
\end{array}
\right),
\end{equation}
where we ignore any possible phases.  
For $|m_S| \ll \mzp$ the eigenvalues are given by the usual seesaw formula,
\begin{equation}
\label{eqn:seesaw}
 \mathcal{M}_{SZ}^{(1)}
=-\frac{2|m_S|^2}{\mzp}  =-\frac{\mzpr^2}{\mzp},  \quad \quad
\mathcal{M}_{SZ}^{(2)} = \mzp.
\end{equation}

The mass of both the $\zp$ gauge-boson and the singlino are governed by $|m_S|$ which is naively of the same order as the other soft scalar masses
$\sim 100$ TeV. However, there is an interesting limit with  $\gz \ll \lambda$ in which the
 fine-tuning required in the Higgs sector leads to smaller values for $|m_S|$.
The singlet's vev, $\svev$, contributes to the mass of
the up-type Higgs as in Eq.(\ref{eqn:higgsMatrix}). The necessary cancellation in Eq.(\ref{eqn:higgsMatrix}) prevents the singlet's vev from becoming too large or it is impossible to tune the up-type Higgs mass to be of the order of the EW scale if $\lambda$ is order unity. The typical value of $m_{H_u}$ is $ \sim
\gz\mzp/4\pi$, a loop-factor below $\mzp$, so we expect $\svev \sim( \gz/\lambda) 
\mzp/4\pi$. But, this implies an even lower scale for $m_S$,
\begin{equation}
|m_S| \sim \frac{\gz^2/\lambda}{4\pi} \mzp \sim (10^{-2}-10^{-3}) \mzp
\end{equation}
We refer to this phenomenon, where  $|m_S|$ is
lighter than expected, as accidental tuning. It is accidental because
it comes about as a result of the fine-tuning in the Higgs sector and the smallness of
the gauge-coupling, $\gz \ll \mathcal{O}(1)$.

This accidental tuning leads to a $\zp$ gauge-boson and singlino much
lighter than expected. The $\zp$ gauge-boson mass   $\mzpr =
\sqrt{2} |m_S|$ can be light enough to be
produced at the LHC. The singlino is even lighter with a mass
$m_{\tilde{S}} = 2|m_S|^2/\mzp \sim 10^{-3}-10^{-6}\mzp$. It may even
be the LSP as we shall demonstrate below with explicit models.

At low energies there will be a single Standard Model-like Higgs
scalar,  while the other linear combination, as well as the charged Higgs and the pseudoscalar,
 is heavy at the $100\TeV$
scale. The Higgs mass is somewhat heavier than the typical prediction
of the MSSM, due to the $\uonep$ $D$ term and the running of the
effective quartic coupling from $\mzp$ down to the electroweak scale.

\section{Model building}\label{sec:model}
\subsection{Charge assignments}
We first outline some general considerations for model building in
this scenario and then present a particular model which satisfies all
of these requirements. Variations on most of these assumptions
are possible, but beyond the scope of this paper.

The free parameters are the $\uonep$ charges of the particles, $\gz$,
$\lambda$, the exotic Yukawa couplings, $\mzp$, and the supersymmetry
breaking scale $\lams$.

We will consider scenarios in which  $\uonep$
is anomaly free under the visible sector fields. This, along with the
need for radiative
breaking, will require the introduction of exotic fields. In
principle, since some of the hidden sector fields
must carry $\uonep$ charges they could also contribute to the anomaly
cancellation. However, such hidden sector fields would have to  be
chiral. If they are to have masses characteristic of the
hidden sector dynamics the $\uonep$ would have to
be broken in the hidden sector. There would therefore be a tendency
for the entire
$\uonep$ supermultiplet to decouple at around the supersymmetry
breaking scale, making
it more difficult to mediate the supersymmetry breaking. We will
therefore assume for simplicity that the hidden sector fields are
non-chiral under $\uonep$.

We will also assume that all of the visible sector fields carry $\uonep$ charges, that there
is a single Standard Model singlet $S$ which not only breaks $\uonep$ but also generates
an effective $\mu$ parameter and exotic masses, and that all Standard Model Yukawa couplings
 are allowed. The latter will include the Dirac coupling for the right-handed neutrino, but
 we will comment on a variation in which this is forbidden.

 With a large $\svev$ there is a danger that the quark and/or slepton fields could become
 tachyonic due to the $\uonep$ $D$ terms, leading to charge and color breaking. We of course
 require that this does not occur.

 Finally, the LSP in these models is usually one of the Standard Model gauginos.
 Because of the well-known difficulties with a bino LSP, we will choose the
 $\uonep$ charges to ensure a wino LSP instead. We do not make any a priori requirements
 concerning gauge unification, exotic decays, kinetic mixing between the $U(1)_Y$ and $\uonep$
 gauge bosons or gauginos, or $R$-parity, but will comment on all of them below.

\subsection{A model}
There are many possible $\uonep$ charge assignments for the ordinary and exotic fields
in a supersymmetric theory~\cite{Erler:2000wu,Langacker:2008}.
The most commonly studied are based on the breaking of the $E_6$ group
to $SU(5) \times U(1) \times U(1)$, which yields an anomaly-free model
consistent with gauge unification~\cite{Hewett:1988xc,Langacker:1998tc,King:2005jy}.
However, it is rather complicated,
involving three $S$-type fields, three pairs $D$ and $D^c$ of exotic charge $\mp 1/3$
quarks, as well as multiple $SU(2)$ doublets which can be interpreted as
extra Higgs doublets or as exotic lepton doublets. The latter ensure a bino LSP
when combined with the $\zp$ mediation scenario.

We will therefore explore an alternative model, characterized by a single $S$ field
and family universal charges.
To ensure a wino LSP we will not introduce any exotic $SU(2)$ doublets
(i.e., no exotic leptons or extra Higgs pairs), but will allow $n_D$ pairs $D, D^c$
of exotic quarks with weak hypercharge $\pm y_D$, and $n_E$ pairs $E, E^c$
of exotic leptons with weak hypercharge $\pm y_E$. The exotics are non-chiral
with respect to the Standard Model gauge group, but chiral with respect to
$\uonep$. Without loss of generality, we can assume family-diagonal exotic
Yukawa couplings
\begin{equation}
\label{eqn:exoticsuperpotential}
W_{\rm exotic}= S\left( \sum_{i=1}^{n_D} y_{D_i} D_i D_i^c + \sum_{j=1}^{n_E} y_{E_j} E_j E_j^c\right)
\end{equation}
(cf. (\ref{eqn:superpotential})). In practice, we will take a common value $y_D$ for each $y_{D_i}$
and similarly for the $y_{E_j}$.

The anomaly conditions are analyzed in  Appendix \ref{app:anomalies}. It is found that the
simplest solution to the mixed anomaly constraints requires $n_D=3$ color triplet pairs
with hypercharge (electric charge) $Y_D=\mp 1/3$, and $n_E=2$ singlet pairs with
$Y_E=\mp 1$. There are two 2-parameter solutions for the $\uonep$ charges,
for which the quark doublet and $H_u$ charges $Q_Q$ and $Q_2$ are free parameters
(after making the normalization $Q_1=1$), and two especially simple 1-parameter special cases in which
$Q_Q$ is fixed. We will mainly but not exclusively consider one of these special cases, with
charges listed in Table \ref{tab:charges}. We reemphasize that this is only a particularly simple example
of a large range of possibilities.

\begin{table}[htdp]
\begin{center}
\begin{tabular}{|c|c|c|c|}
\hline
\hspace{2mm}$H_d$\hspace{2mm} &    $1$                &\hspace{2mm} $L$ \hspace{2mm}& \hspace{1mm} $\frac{2}{3}-\frac{1}{3}x$ \hspace{1mm} \\
$H_u$ & $x$                   & $e^+$   & $-\frac{5}{3}+\frac{1}{3}x$ \\
$S$      &  $-(1+x)$         & $N^c$   & $-\frac{2}{3}(1+x)$ \\
$Q$      &  $-\frac{1}{3}$  & $D$       & $\frac{8}{9}+\frac{2}{9}x$ \\
$u^c$   &$\frac{1}{3}-x$ & $D^c$  & $\frac{1}{9} +\frac{7}{9}x$ \\
$d^c$   & $-\frac{2}{3}$  & $E$      & $\frac{5}{3}-\frac{1}{3}x$ \\
              &                           & $E^c$ & $-\frac{2}{3} +\frac{4}{3}x $ \\
\hline
\end{tabular}
\end{center}
\caption{$\uonep$ charges for a particular anomaly free model. We assume $Q_{H_u}\equiv Q_2
\equiv x\ne -1$, and usually take $x=-1/4$ in our numerical examples.}
\label{tab:charges}
\end{table}

\section{Phenomenology}
\label{sec:pheno}

The low energy phenomenology depends on several free parameters, such
as the charge assignments, the exotics' Yukawa
couplings, the PQ-symmetry breaking coupling $\lambda$, and the
$\uonep$ gauge coupling $\gz$. In this section we  explore
the parameter space spanned by these choices and arrive at a global
picture of the low-energy phenomenology.

\begin{figure}
\begin{center}
\includegraphics[scale=0.35]{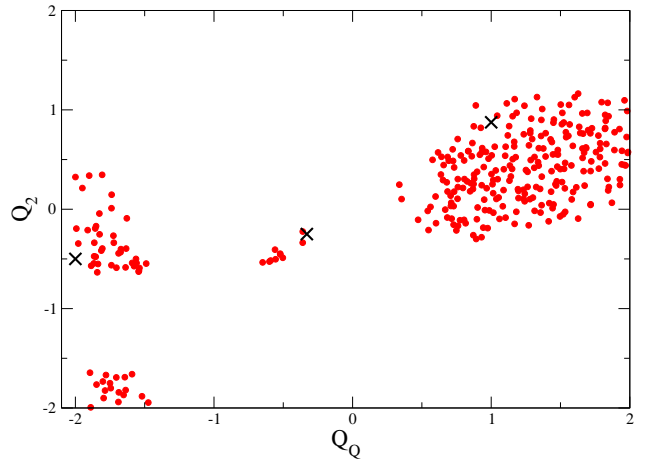}
\end{center}
\caption{The red dots in the $Q_Q-Q_2$ plane represent points for which a
  viable solution exists and where the electroweak scale is obtained,
  using the ``$+$'' solution in (\ref{cubicsoln}). We
  fixed $\lambda = y_D = 0.5$ and $y_E = 0.1$. The regions
  of viable solutions will change as we vary these parameters,
  but we have verified  that the overall structure remains unchanged. We
  then picked three representative points, one from each ``island''
  (indicated by ``x''),  and investigated the resulting spectrum
  in detail. The point at $Q_Q=-1/3$ corresponds to the charges in
  Table  \ref{tab:charges} with $x=-1/4$.} 
\label{fig:Charges-scan}
\end{figure}

We begin with the charge assignment. After imposing the anomaly
cancellation conditions 
and normalizing the down-type Higgs
$\uonep$ charge to unity, $Q_1 = 1$, one is left with two
undetermined charges, namely, $Q_2$ and $Q_Q$ (the up-type Higgs
charge and the left-handed quarks' charge, respectively). 
The other charges are given by the equations in Appendix \ref{app:anomalies}.
The up-type
Higgs charge, $Q_2$, has to be small or otherwise it is either
impossible to turn $m_S^2$ negative or fine-tune $m_{H_u}^2$ to be
small. In Fig. \ref{fig:Charges-scan} we present a scan of the points
in the $(Q_Q,Q_2)$ space where a solution is possible, i.e., where it is
possible to obtain the EW scale without driving any of the scalars
tachyonic or have any other supersymmetry partner too light.
The scan utilizes the
 the ``$+$'' solution in (\ref{cubicsoln}), but we have verified that the ``$-$'' solution is similar.

One simple choice of charges has $Q_2 = -1/4$ and $Q_Q = -1/3$
(see Table \ref{tab:charges}). We normalize the coupling $\gz$ to the hypercharge $U_Y(1)$ at the cutoff $\lams$,
\begin{equation}
\label{eqn:normalization}
\gz^2 = {\cal N}^2 g_Y^2 \frac{Tr~Y_i^2}{Tr~Q_i^2}.
\end{equation}
We leave ourselves the freedom to choose a factor $ {\cal N}^2$ of order unity.
With the above choice of charges, $\gz \sim \frac{2}{3} g_Y  {\cal N}\sim 0.23  {\cal N}$ at the SUSY breaking scale. 

The other important parameters  are the colored
exotics' Yukawa, $y_D$, and $\lambda$. To gain a better
insight into the range of possibilities in this class of models we
performed a scan over both parameters and demanded that the EW scale
is obtained by fine-tuning the SUSY breaking scale $\Lambda_S$. The
details of this procedure are summarized in Appendix~\ref{app:RGEs} .

\begin{figure}
\begin{center}
\includegraphics[scale=.34]{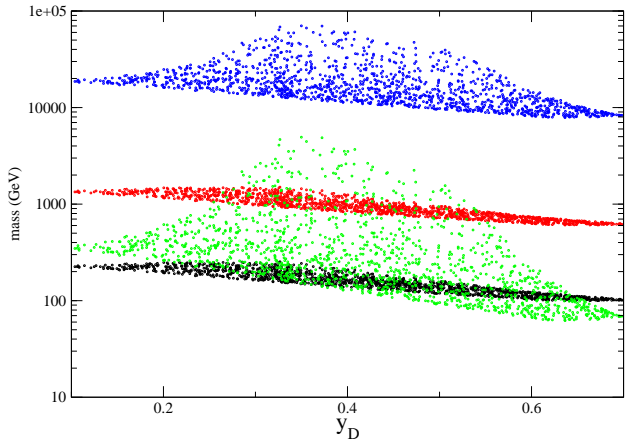}
\end{center}
\caption{A plot of the low energy masses as a function of the colored exotic fields' Yukawa coupling, $y_D$. $\zp$ gauge-boson (Blue, first from top right), gluino (Red, second from top right), wino (Black, third from top right) and singlino (Green, fourth from top right). This spectrum corresponds to the charge assignment $Q_Q=-1/3$ and $Q_2=-1/4$. The $\uonep$ gauge-coupling is set according to Eq.(\ref{eqn:normalization}) with a factor of $ {\cal N}=0.5$. The bino mass is slightly lighter than the gluino mass and is not shown  in order to reduce clutter. The spread corresponds to a variation in $\lambda$, the Higgs coupling to the singlet.}
\label{fig:Spectrum-Q1}
\end{figure}

As commented in Section~\ref{sec:generic},  we expect the $SU(3)_C
\times SU(2)_L \times U(1)_Y$ gauginos to be light. In our particular
model, where none of the exotics are charged under the SM $SU(2)$, the
lightest gaugino is typically the wino. In addition, the singlino is
usually also quite light as a consequence of the seesaw mass relations
of Eq. (\ref{eqn:seesaw}). In Fig. \ref{fig:Spectrum-Q1}, we plot the
low-energy spectrum as a function of the colored exotics' Yukawa
coupling, $y_D$.  

\begin{figure}
\begin{center}
\includegraphics[scale=.34]{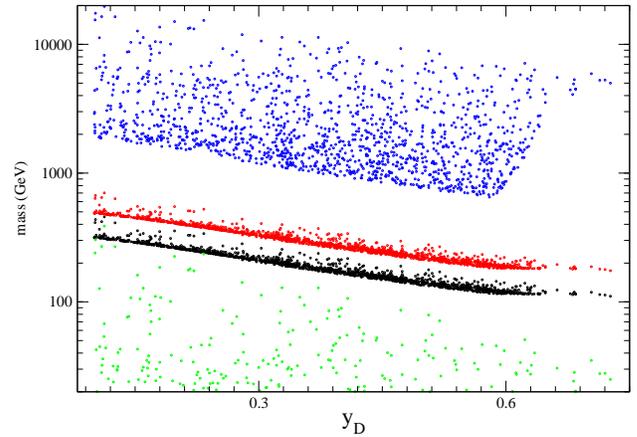}
\end{center}
\caption{Same as Figure \ref{fig:Spectrum-Q1}, except $Q_Q = -2$ and $Q_2 = -1/2$. The coloring and ordering from top right remains the same. 
The bino in this case is actually heavier than the gluino.}
\label{fig:Spectrum-Q2}
\end{figure}

\begin{figure}
\begin{center}
\includegraphics[scale=.34]{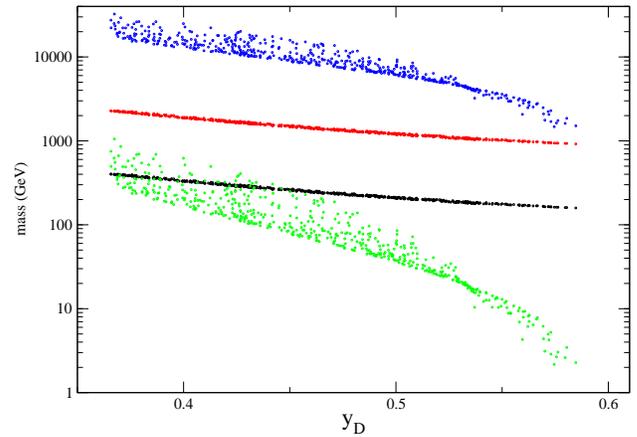}
\end{center}
\caption{Same as Figure \ref{fig:Spectrum-Q1}, except $Q_Q = 1$ and $Q_2 = 7/8$. The coloring and ordering from top right remains the same.}
\label{fig:Spectrum-Q3}
\end{figure}

It is important to consider the other islands of acceptable charge assignments shown in Fig. \ref{fig:Charges-scan}. As we
already saw, there is not much freedom in the choice of $Q_2$. Choosing it too large and it becomes impossible to break
$\uonep$. However, we can try to choose a larger value for
$Q_Q$. To pick a point on the left island we take $Q_Q = -2$ and $Q_2
= -1/2$ \footnote{If one requires rational charges one needs $Q_2 = -79/160$}. From the island on the
right we take $Q_Q=1$ and $Q_2 = 7/8$. We verified that the
conclusions which follow hold for other choices and are 
generic. The resulting spectra are shown in
Figs. \ref{fig:Spectrum-Q2} and \ref{fig:Spectrum-Q3}, respectively.

It is clear from the figures that the low-energy spectrum has a
variety of patterns in the space of $\zp$-mediated supersymmetry
breaking models. In particular, different 
ordering of the MSSM gauginos and the singlino could give rise to very
different phenomenology, and the appearance of a light $\zp$
gauge-boson may prove to be the strongest indicator of the nature of  the
SUSY breaking mechanism. In Table \ref{benchmark} we give six
benchmark points illustrating the possible variations in low-energy
parameters for different charge choices and couplings. Point 6, which is the last column in the table, has to be interpreted carefully. By itself this spectrum is inconsistent since such a large Supersymmetry breaking scale, $\lams=6\times10^{11}\GeV$ will induce gaugino masses much larger than the electroweak scale through gravity mediation. This conclusion may be evaded if some form of sequestering takes place, but we will not attempt such a construction here.

\begin{table}
\begin{tabular}{|c|c|c|c|c|c|c|}
\hline
& 1 & 2 &  3 & 4  & 5 & 6\\
\hline $Q_2$ & $-\frac{1}{4}$& $-\frac{1}{4}$ & $-\frac{1}{4}$  &  $-\frac{1}{2}$&
$-\frac{1}{2}$ & $\frac{7}{8}$\\ 
 $Q_Q$ & $-\frac{1}{3} $& $-\frac{1}{3}$ &  $-\frac{1}{3}$ & $-2$ &  $-2$ & $1$\\
$\gz$ & 0.45  & 0.23 & 0.23  & 0.06& 0.04 & 0.055\\ 
$\lambda$ & 0.5 & 0.8 & 0.8 & 0.3 & 0.3 & 0.6\\ 
$y_D$ & 0.6 &0.7 & 0.8 & 0.4 & 0.6 & 0.55\\ 
$y_E$ & 0.6& 0.6 & 0.6 & 0.1 & 0.1 & 0.1\\
\hline
$\lams$ & $5\times10^{10}$ & $9\times10^{10}$ & $4\times10^{10}$ & $3\times10^9$ & $5\times10^8$ & $6\times10^{11}$ \\
$\svev$ & $2\times 10^5$ & $7 \times 10^4 $ & $6 \times 10^4$ & $2
\times 10^5$ & $ 8\times 10^4$ & $2\times 10^4$\\ 
$\tan\beta$ & 20 &29 & 33 &45 & 60 & 23\\ 
$M_1$ & 2700 & 735 & 650 & 760 & 270 & 185\\ 
$M_2$ & 710 & 195 & 180 & 340 & 123 & 178\\ 
$M_3$ & 4300 & 1200 & 1100 &540 & 200 & 1040\\ 
$m_H$ & 140  & 140  & 140 &140 & 140 & 140\\ 
$m_{\tilde Q_3}$ &$1 \times 10^5$ & $ 5 \times 10^4 $ & $4 \times
10^4$ & $ 8\times 10^4$& 
$4\times 10^4$ & $4\times 10^4$ \\  
$m_{\tilde L_3}$ &$ 3 \times 10^5$ & $10^5$ &$10^5$ & $2 \times10^4$ &
$10^5$ & $1.2 \times 10^5$ \\ 
$m_{3/2}$ &  890 & 3600 & 810 & 3 & 0.1 & $10^5$\\ 
$m_{\tilde{S}}$ & 4300 & 230 & 160 & 31 & 4 & 11\\ 
$M_{Z^\prime}$ & $ 7 \times 10^4$& $1.5 \times 10^4$& $1.3 \times 10^4$ &
5600 & 2100 & 3400\\ 
\hline
\end{tabular}
\caption{Model inputs and superpartner spectra of six representative models. The masses are in GeV.  We  fix $\mzp = 10^6$ GeV. The masses of the first two generations of squarks and sfermions are typically larger than that of the third. The input parameters $\lambda$, $\gz$ and $y_{D,E}$ are defined
at $\Lambda_S$. The spectra are calculated using full Renormalization
Group Equations (RGE)~\cite{Yamada:1993ga,Martin:1993zk,Yamada:1994id,Luo:2002ey,Kielanowski:2003jg}. There is a
theoretical uncertainty due to multiple RGE thresholds which mainly
affects $m_H$, leading to a several GeV uncertainty.  The gravitino
mass is calculated by $m_{3/2} = \Lambda_S^2/M_P$,
where $M_P$ is the reduced Planck mass,  assuming $\Lambda_S
\sim \sqrt{F}$. 
There could be deviations from this relation in some
SUSY breaking models which could lead to a gravitino mass that is
different by up to a couple orders of magnitude (typically lower).}
\label{benchmark}
\end{table} 

\subsection{LHC phenomenology}

\subsubsection{Gluino}

\begin{figure}
\begin{center}
\includegraphics[scale=.6]{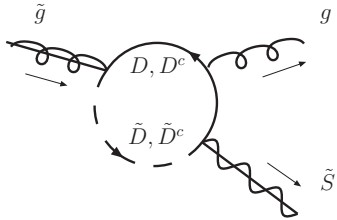}
\end{center}
\caption{The gluino can decay through colored exotic states into the
  singlino and a gluon. The other diagram in which the gluon is attached
  to the scalar propagator is suppressed. This decay channel can
  compete with the more standard decay of the gluino through off-shell
  squarks.} 
\label{fig:gluinoDecay}
\end{figure}

Since the colored scalars are all very heavy, the LHC will
predominantly produce gluino pairs.  The gluinos will consequently
decay either through a 3-body 
off-shell squark ($\tilde{g}\rightarrow q~\bar{q}~\tilde{\chi}_i$,
where $\tilde{\chi}_i$ is one of the gauginos) or a 2-body loop
induced process. The 3-body decay usually dominates,  leading to a gluino
life-time
\begin{equation}
\label{eqn:gluino-lifetime}
\tau_3 = 4\times 10^{-16} \mbox{sec}~ \left(
\frac{m_{\tilde{Q}}}{10^2\TeV}\right)^4
\left(\frac{1\TeV}{M_3}\right)^5 ~\propto~ \frac{1}{\gz^6} . 
\end{equation}
Interestingly enough, the decay induced by the loop with exotic
matter, shown in Fig. \ref{fig:gluinoDecay}, can be the leading
effect. We will discuss here mainly the processes which result in a
singlino, as it usually has the largest coupling to the exotics. Two
body decays into other MSSM gaugino states will be somewhat suppressed
(although they could be important in certain cases) and the Higgsino
does not couple to the exotic sector directly.  The expression for the
decay width of $\tilde{g}\rightarrow \widetilde{S}~g$ can be extracted
from \cite{Toharia:2005gm} with the appropriate changes.  The gluino
decay corresponds to the dimension-5 operator
$\bar{\widetilde{S}}\sigma^{\mu\nu}\gamma_5\tilde{g}^aG_{\mu\nu}^a$,
where the presence of $\gamma_5$ is due to the Majorana nature of the
gluino and singlino. As pointed out in \cite{ArkaniHamed:2004yi}, this
operator is P (and C)-odd and therefore must vanish in the limit where
the left and right-handed heavy scalars are degenerate. Indeed, the
decay width for this channel is given by,
\begin{eqnarray}
\Gamma_{\tilde{g}\rightarrow \widetilde{S}~g} &=& \frac{1}{8\pi}\frac{2
  g_S^4}{\left(32\pi^2\right)^2}\left(\frac{M_3^2-m_{\tilde{S}}^2}{M_3}\right)^3
\\\nonumber &\times& n_D^2 m_D^2 y_D^2 \left(C_0^L-C_0^R\right)^2,
\end{eqnarray}
where $C_0^{L,R}$ are the Passarino-Veltman functions
\cite{Passarino:1978jh},  involving the left (right)-handed exotic
scalars. They are given by (neglecting terms which are suppressed by
ratios of the gaugino masses to the exotic matter mass), 
\begin{equation}
\label{eqn:PVfunc}
C_0^{L,R} = \frac{m_{\tilde{D}_{L,R}}^2-m_D^2+m_{\tilde{D}_{L,R}}^2\log\left(m_D^2/m_{\tilde{D}_{L,R}}^2\right)}{\left(m_{\tilde{D}_{L,R}}^2-m_D^2\right)^2},
\end{equation}
where $\widetilde{D}_L$ ($\widetilde{D}_R$) is the scalar component of $D$
($D^c$). Since these fields are chiral under $\uonep$ they evolve
differently under the RGE running and can differ significantly in
mass. Parametrically, the 2-body channel leads to a life-time (assuming no phase space suppression), 
\begin{equation}
\tau_2 \approx \frac{8}{n_D^2}
10^{-18}\mbox{sec}\left(\frac{m_D}{10^2\TeV}\right)^2 \left(\frac{1\TeV}{M_3}
\right)^3,
\end{equation}
The exact value could be longer or shorter depending on the precise
value of $C_0^{L,R}$. This
analysis shows that it is potentially competitive with the standard
3-body mode and can lead to an interesting exotic decay of the
gluino. 
In Table \ref{tbl:gluinoLT} we contrast the life-time
associated with the exotic 2-body mode versus the standard 3-body
channel for the different benchmark points considered above. 
The relative branching ratio is very sensitive to the detailed
model  parameters. This is to be expected since the two-body width
depends sensitively both on the mass splitting of left and right
handed exotic scalars, as well as the mass of the exotic
fermions. These quantities are in turn determined by charge
assignments and exotic Yukawa couplings. 

We remark here that the 2-body decay could give rise to very
interesting collider signals if the singlino is not the LSP and decays
subsequently (more on singlino decay in the next section).

\begin{table}
\begin{tabular}{|c|c|c|c|c|c|c|}
\hline
& 1 & 2 &  3 & 4  & 5 & 6\\
\hline 
$\tau_2$ &  $9\cdot 10^{-13}$ & $8 \cdot 10^{-19}$ & $6 \cdot
10^{-19}$ & $6 \cdot 10^{-15}$
& $5 \cdot 10^{-14}$ & $3\cdot 10^{-18}$\\  
$\tau_3$ &  $4 \cdot 10^{-19}$ & $7 \cdot 10^{-18}$ & $7 \cdot
10^{-18}$ & $ 10^{-16}$
& $10^{-15}$ & $8 \cdot 10^{-18}$\\  
\hline
\end{tabular}
\caption{The gluino life-time  (sec) for the 2-body channel versus the 3-body
  mode for the different benchmark points presented in Table
  \ref{benchmark}.} 
\label{tbl:gluinoLT}
\end{table}

It is interesting to compare the gluino decay signature in our case
with that of the split supersymmetry scenario. In split SUSY,  the
gluinos will also  decay either through a 3-body 
off-shell squark ($\tilde{g}\rightarrow q~\bar{q}~\tilde{\chi}_i$,
where $\tilde{\chi}_i$ is one of the gauginos) or a 2-body loop
involving both the squarks and the Standard Model fermions
\cite{Toharia:2005gm,Gambino:2005eh}. A log enhancement of the 2-body
channel  associated with the third generation squark-quark loop, as well
as the mixing between the LSP and the Higgsino,  are
important for the two-body decay to be comparable with the 3-body.
Since in our case the Higgsinos are both very heavy, there is no such
log enhancement and two body decays are dominated by the exotic
loop. Given that the  two-body decay  is induced by
completely different virtual 
states, we expect the resulting branching ratio
of $\tilde{g}
\rightarrow g \widetilde{S}$ 
will be quite different from that of
split supersymmetry.  For example, for the squark masses in our
scenario, the gluino life time 
is always too short to produce sizable displaced vertices. In the
split supersymmetry scenario considered in
\cite{Toharia:2005gm,Gambino:2005eh}, the three body channel always
dominates over the two body one within the same range of squark masses,
while the situation could be very different in our
scenario. 

\subsubsection{The LSP and other Inos}

In general, the pattern of MSSM gaugino masses depends on both the
charge assignments and the choice of the exotic sector. As a result of
the absence of exotic doublets, which is  a specific choice we made
here, the wino is the lightest MSSM gaugino. 
The mass of the bino in our model is comparable to the gluino's and
never serves as the LSP. The light Inos include the wino, the
singlino and possibly the gravitino. As shown in the previous section
and illustrated in Figs. \ref{fig:Spectrum-Q1}-\ref{fig:Spectrum-Q3},
this model may admit different orderings of the light Inos, and we
discuss the different possibilities below. 

The mass of the gravitino does not affect LHC
phenomenology in this model. If the gravitino is not the LSP, it will
not be produced at the LHC. At the same time, if it is the LSP, the
range of gravitino mass  implies that the NLSP will decay
outside the detector. Due to the decoupling of the scalars, the NLSP 
 is neutral. Therefore, such decays will not be observable
at the LHC. 

The case of a singlino LSP with decoupled electroweak gauginos does
not produce observable 
effects at the LHC either. In this case, the only way to
produce the singlino is through the decay of the  $Z^{\prime}$. However, the decay
mode will be dominated by $Z^{\prime} \rightarrow \widetilde{S}
\widetilde{S}$, which is again not observable. 

There are several more interesting scenarios with either wino LSP or
NLSP. \\

\noindent\underline{\it Wino LSP only} \\

At tree level, the neutral and charged winos are degenerate, and the
mass splitting induced by mixing with the Higgsinos is negligible for
the large effective $\mu$ of this scenario. However, there is an
important one loop radiative correction which increases the charged
wino mass by $\sim 160\MeV$ with respect to the $\widetilde W^0$ state
\cite{Gherghetta:1999sw,Pierce:1996zz,Chen:1996ap,Feng:1999fu,Chattopadhyay:2006xb,Ibe:2006de}.
This allows for the decay $\widetilde W^+ \rightarrow \widetilde W^0 +
\pi^+$ with a lifetime around $1.4 \times 10^{-10}$ sec, corresponding
to a track length and displaced vertex around 4 cm from the production
vertex in a detector, as has been studied extensively in connection with anomaly mediation~\cite{Randall:1998uk,Giudice:1998xp}. \\

\noindent\underline{\it Wino NLSP and Singlino LSP} 

The wino can only decay to the singlino by mixing through the Higgsinos, leading to  a suppression of the decay width. If there is no further phase-space suppression then the life-time for $\widetilde{W} \rightarrow h + \widetilde{S}$ is approximately,
\begin{eqnarray}
\label{eqn:winoLT}
\tau &\sim& \frac{4 \pi}{g_W^2}  \left( \frac{\svev \tan\beta}{v}\right)^2 M^{-1}_{\tilde{W}} \nonumber \\
&\sim& 10^{-17} \mbox{sec} \left(\frac{100\GeV}{M_{\tilde{W}}}\right),
\end{eqnarray}
where the ratio of the singlet's VEV to the electroweak scale stems from the Higgsino-singlino mixing.
Of course, the lifetime would be longer if there is phase space suppression or the decay is via
a virtual Higgs, and it is even possible in that case that there would be a displaced vertex.  
\\

\noindent\underline{\it Singlino NLSP and Wino LSP} \\

The singlino decay into wino has a similar life-time to the reversed
process (with $M_{\tilde{W}}$ replaced by $m_{\tilde{S}}$ in
Eq.(\ref{eqn:winoLT})). The singlino could be produced through
$Z^{\prime}$ decay so this channel is potentially interesting and
should be investigated further. \\

\subsubsection{$\zp$ production and decay}

\begin{figure}
\begin{center}
\includegraphics[scale=.3]{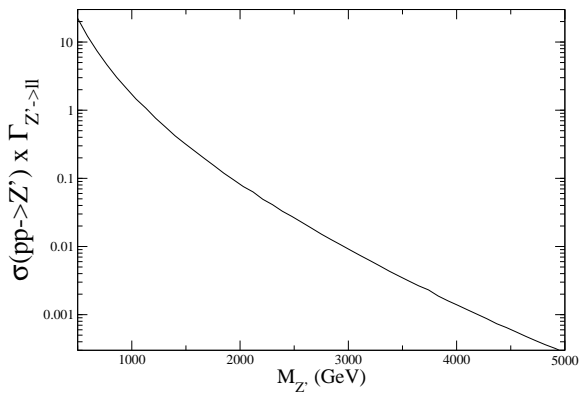}
\end{center}
\caption{A plot of the $\zp$ gauge-boson production cross-section
  times the leptonic branching ratio, which includes
  both $\mu^+ \mu^-$ and $e^+ e^-$ final states. The $U(1)'$ charge
  assignment used in generating this plot is $Q_2=-1/2$ and $Q_Q=-2$ and the coupling was chosen nominally to be $\gz = 0.06$.}
\label{fig:zprime-prodbr}
\end{figure}
In some of the benchmarks we presented, the $\zp$ is light enough to
be produced at the LHC. This happens when $|Q_Q| \gtrsim 1$ and corresponds
to the islands on the left and right in Fig.\ref{fig:Charges-scan}. In
this case the normalization of $\gz$ becomes important 
since $Tr~Q_i^2$ is larger, and $\gz \approx \frac{1}{10} g_Y$ and
hence considerably smaller. This would normally be harmful, 
causing the wino to be too light. However, this is avoided here because
the wino RGE has a term  proportional to $Q_Q^2$ (and the other doublets' charges) and
therefore the wino receives a large contribution as well. Together with the accidental tuning discussed
above it is possible and even likely to have the $\zp$ gauge-boson in
the observable spectrum as well as a very light singlino.  

To have a light $\zp$ gauge boson that is
accessible at the LHC, typically requires a smaller gauge coupling
$\gz$. With a fixed spontaneous symmetry
breaking scale,  such a choice actually results in enhanced discovery
potential at the LHC. Although the parton level total cross section is
proportional to $\gz^2$, the parton distribution function depends  inversely
on a large power of $m_{\zp} \propto \gz$.

In Fig. \ref{fig:zprime-prodbr} we plot the $\zp$ production
cross-section times the leptonic branching ratio. If the $\zp$ is not
too heavy, $\mzpr < 4-5\TeV$ it will likely be an easy task to observe
this resonance and determine its mass through its leptonic decay.
Once its existence is established, it may be possible to uncover other
and more difficult decay channels, such as $\zp \rightarrow
\widetilde{S}\widetilde{S}$ etc. (for the possible utilization of a $\zp$ in
disentangling more difficult channels see
Ref. \cite{Baumgart:2006pa,Langacker:2008}). A full discussion of the discovery reach
and experimental challenges is beyond the scope of this paper, and we
leave it for a future and more comprehensive study.  

\subsubsection{Higgs mass}

At low energies there remains one light Higgs in the spectrum. Its mass is given as usual by $m_H^2 =2 \lambda_H v^2$, where $v = 174\GeV$ and $\lambda_H$ is the quartic coupling. 
The value of $\lambda_H$ at low energies is determined by matching it to the supersymmetric contribution at $\mzp$ and running it down to the electroweak scale,
\begin{eqnarray}
16\pi^2 \frac{d \lambda_H}{dt} &=& 12\left(\lambda_H^2+\lambda_Hy_t^2- y_t^4 \right) \\\nonumber
\lambda_H(\mu \approx \mzp) &=& \frac{1}{4}(g_2^2 + g_Y^2) + \gz^2 Q_2^2 +\frac12 \lambda^2 \sin^22\beta.
\end{eqnarray}
The $F$-term contribution to the quartic, $\lambda^2 \sin^22\beta$, is negligible since $\tan \beta \gg 1$. The $D$-term contribution from the $\uonep$ vector multilplet, $\gz^2 Q_2^2$, is usually smaller than the $SU(2)\times U(1)_Y$ D-term because both $\gz$ and $Q_2$ are not very large.

This leads to a prediction of the Higgs mass which is insensitive to the precise details of the high-energy parameters. It is predominantly affected by the running from $\mzp$ down to the electroweak scale and yields,
\begin{equation}
 m_H = 140\GeV
\end{equation}
with an uncertainty of a few percent coming from the precise matching and the value of $\mzp$ (which we fixed at $\mzp=1000\TeV$ for concreteness).

\subsection{Cosmology}
\subsubsection{The Wino}
We have deliberately chosen the $U(1)'$ charges and exotics in our example construction to avoid a bino LSP. This is because the bino lacks any efficient annihilation or co-annihilation mechanism for the large scalar masses and effective $\mu$ parameter favored in the scenario, leading to too much cold dark matter (CDM). (For a recent discussion, see, e.g., \cite{Arkani-Hamed:2006mb}.) On the other hand, a wino LSP and its nearly degenerate charged partner, which have been studied extensively, especially in connection with anomaly mediated models \cite{Randall:1998uk,Giudice:1998xp}, can annihilate efficiently into gauge bosons. In fact, for pure thermal production the CDM density is too low for the several hundred GeV mass range we have assumed, yielding \cite{Arkani-Hamed:2006mb} \begin{equation} \Omega h^2 \sim 0.021 \left( \frac{M_2}{1\ {\rm TeV}} \right)^2, \end{equation} compared to the observed value $0.111 \pm 0.006$ from WMAP and galaxy surveys \cite{Spergel:2006hy,Yao:2006px}. However, the CDM density can be considerably larger for non-standard cosmological scenarios 
\cite{Gherghetta:1999sw,Moroi:1999zb,Salati:2002md,Rosati:2003yw,Profumo:2003hq,Giudice:2000ex,Bourjaily:2005ax,Chung:2007vz}.

\subsubsection{The Gravitino}
\label{sec:gravitino}

\begin{figure}
\begin{center}
\includegraphics[scale=.34]{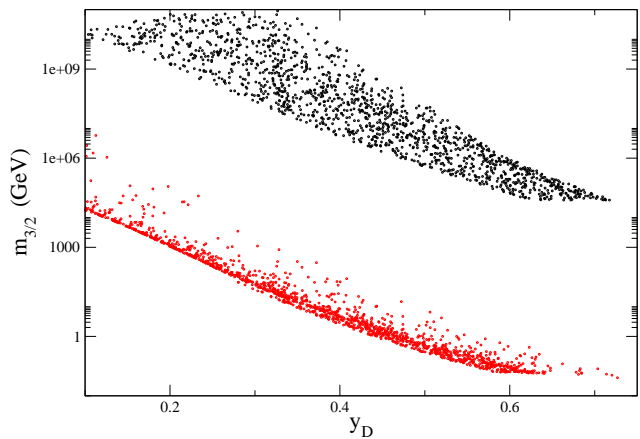}
\end{center}
\caption{A plot of the gravitino mass, $m_{3/2}$ as a function of the colored exotics' Yukawa coupling. The black (upper) points correspond to the charge assignment with  $Q_Q = -1/3$ and $Q_2=-1/4$, whereas the red (lower) points correspond to the assignment $Q_Q =-2$ and $Q_2=-1/2$. The high value of the black points to the left are because a large $\Lambda_S$ is needed to compensate the small Yukawa for those charges.}
\label{fig:gravitinoMass}
\end{figure}

Another particle of interest to low-energy phenomena is the gravitino,
with a mass given by
\begin{equation}
m_{3/2} \sim \frac{F}{k \sqrt{3} M_P} \sim \frac{2.4 \ {\rm eV}}{k}\
\left( \frac{\sqrt{F}}{100 \ {\rm TeV}} \right)^2,
\end{equation}
where $M_P = 2.4 \times 10^{18}$ GeV is the reduced Planck mass
and $k$ depends on the details of the supersymmetry breaking mechanism
in the hidden sector, but is typically $\le 1$. We will take
$\sqrt F = \Lambda_S$ and $k=1$. The gravitino then depends very strongly
on the SUSY breaking scale. The value of $\Lambda_S$ does affect the
other masses, because it 
determines the overall scale separation (recall that we tune
$\Lambda_S$ to obtain the EW scale while keeping $\mzp$ fixed),
but the dependence is only logarithmic. The 
symmetry breaking pattern in our model depends only logarithmically on
the supersymmetry breaking scale. Therefore, the gravitino mass is 
exponentially sensitive to the choice of the charges and
couplings, as shown in Fig. \ref{fig:gravitinoMass}, and it may provide a sharp discriminator in the model
space. 

In this section, we will focus on cosmological implications and
constraints on the gravitino mass  (A good summary is given in 
\cite{Giudice:1998bp}.). However, one should keep in mind that these
constraints are fairly indirect and can be overcome as mentioned
below. 

A stable (LSP) gravitino could overclose the universe unless it is
lighter than a few keV (as in normal gauge mediation). However, this
difficulty could be evaded if the reheating temperature $T_R$ after
inflation is rather low (i.e., $T_R \lesssim 10^8 m_{3/2}$). 

The strongest constraint on gravitino mass comes from its interactions
with other superpartners present in the early 
universe. Exact constraints on the parameter space are quite sensitive to the particle spectrum and interactions. A detailed study based on
the spectrum, which is quite unique, is beyond the scope of this paper. In the following, we will very briefly summarize the results
from early studies, see, for example,  \cite{Feng:2004mt} and \cite{Kohri:2005wn}, comment on the relevance to our scenario, and point out cases where more careful studies need to be done.

Decay processes involving the gravitino are typically constrained by big
bang nucleosynthesis (BBN), due to its long life-time. If the decay products involve
hadrons, any such decay with a lifetime longer than 1 second is
strongly constrained by deuterium and helium abundances. On the other
hand, if the 
decay process only induces electromagnetic showers then only a
lifetime longer than $10^4$ seconds is strongly constrained. 

We begin with the case in which the gravitino is not the LSP. Such gravitinos will be produced during reheating. The gravitino decay
into the LSP may lead to unacceptable modifications of BBN if there is any significant component of hadrons  in the
decay. Such effects have been studied carefully, e.g., in \cite{Kohri:2005wn}. One way around this is for the gravitino lifetime
to be shorter than $\sim$ 1 s \footnote{If the decay products only contain photons, BBN constraints could be easily satisfied if the life-time is
less than $10^4$ sec, which corresponds to a gravitino mass of about 1
TeV. However, in our examples, we always have a light wino. Therefore, 
$\widetilde{G} \rightarrow W/Z + \widetilde{W}$ will usually lead to hadrons.}. 
This typically requires $m_{3/2} \gtrsim 10$
TeV  \footnote{This requires that the gravity mediation effects are
  sequestered. In this case one would again have contributions 
from the anomaly mediation~\cite{Randall:1998uk,Giudice:1998xp} to the
Standard Model gauginos comparable to those of the two-loop 
$Z'$ mediation, while the anomaly mediation would be irrelevant for
the other soft parameters. This hybrid scenario could also use the
mechanism of 
\cite{Moroi:1999zb} to increase the CDM density due to the gravitino
decay into the LSP wino.}. 
Alternatively, the BBN constraints can be satisfied  for a relatively low reheating
temperature, $T_R< 10^6-10^7\GeV$, suppressing the gravitino production.
The only difference in our case from those
well studied scenarios is the decoupled sfermions. It is expected to
affect more significantly the case where the gravitino mass is heavier
than the gluino mass. The enhanced branching ratio of this channel in
the absence of sfermions makes the constraint on the reheating
temperature slightly stronger 
\cite{Kohri:2005wn}.

Alternatively, the gravitino could be the LSP. In this case the
analysis becomes more complicated since the constraints depend on 
the identity of the NLSP. A scenario with a wino as the NLSP is
similar to one with a bino although the numerics are different because
of the smaller branching ratio into photons and the larger
annihilation cross-section during freeze out
\cite{Cyburt:2002uv,Feng:2004mt}. (For the bino case one requires 
the gravitino to be lighter than about 100 MeV.)
A singlino as the NLSP is even less
favorable because its decay into the gravitino must involve mixing
with the Higgsino states, which leads to a suppression of the decay
width. Furthermore, the life-time is very sensitive to the precise
value of $\sqrt{F}$, which we do not have a precise prediction
for. All together the lifetime is generically much longer than a
second regardless of the precise decay mode.  It seems that having a
singlino as the NLSP with appreciable density is pretty much ruled
out. The singlino is expected to be produced in the 
thermal soup since it couples to the  $Z^{\prime}$. Therefore, it is
hard to see how to make this case viable, without resorting to more
exotic cosmologies with large late-time entropy production, such as 
thermal inflation \cite{thermal-inflation1,thermal-inflation2,thermal-inflation3}.

\section{Comments and Alternatives}

\subsection{Other possibilities of $\zp$-mediation}

In this paper, we have focused on a particular scenario of
$\zp$-mediation. Motivated by solving the $\mu$-problem, we have
considered  a singlet-extended MSSM with a PQ-like $\zp$. More
generally, there are of course many other possibilities of $\zp$ which
can play the role of the mediator of supersymmetry breaking, such
as $B-L$, or any other well studied or yet unknown exotic
$U(1)$. As we have demonstrated in the examples presented in this
paper, the detailed spectrum from $\zp$-mediation depends quite
sensitively on the choice of model. However, we would like to
emphasize that the sizable hierarchy between the scalars and the
electroweak-inos will be a very generic feature of the $\zp$
mediation. 

It is of course possible to combine other mediation mechanisms
with the $\zp$ mediation. In those scenarios, we generically expect
that the $\zp$ mediation contribution to the electroweak-ino masses
will be negligible, while the contribution to the scalar masses will
be significant. The challenge of such scenarios is to give plausible
reasons to why some other mediation mechanism will give comparable
contributions as the $\zp$-mediation. Recently, one scenario of such a
combination with anomaly mediation and a hypercharge mediation has
been studied \cite{Dermisek:2007qi}, and a combination with $D$-term
mediation in \cite{Nakayama:2007je}. Further studies on other
possibilities for combining $\zp$-mediation with other mechanisms are
certainly interesting and worth pursuing. 
 
\subsection{An Alternative Model of Neutrino Masses}\label{section:alt}
 $\uonep$ models usually do not allow the large Majorana masses
necessary for the canonical seesaw model 
\cite{Kang:2004ix}. The specific model constructed in Section
\ref{sec:model} allows Dirac masses by assumption, which would have to
be made small by 
fine-tuning. However, in a simple variant~\footnote{The anomaly
  conditions in this variant require 4 singlets $S$ and the $\nu^c$
  charge inferred from (\ref{eqn:nuvariant}). The other conditions are
  unchanged except for the form of the discriminant in
  (\ref{cubicsoln}), which we will not display. The variant
  discriminant vanishes for $Q_Q=Q_2/3$ or for $Q_Q=-Q_1-2Q_2/3$.},
the $\uonep$ symmetry forbids Dirac Yukawa couplings $ y_\nu  {H}_u
{L} {\nu}^c$ at the renormalizable level, but allows them to be
generated by a higher-dimensional operator,  
 \begin{equation}
\label{eqn:nuvariant}
W_\nu=c_\nu \frac{S}{M_P} {H}_u {L} {\nu}^c.
\end{equation}
This naturally yields small Dirac neutrino masses of order $(0.01
c_\nu)$ eV for $\mathcal{S} = 100$ TeV, 
in accordance with observation. (This mechanism has been studied previously
in a more general context~\cite{Langacker:1998ut}.) One cannot say more about the
hierarchy of neutrino masses or mixings without additional assumptions.

\subsection{Exotics and $R$ parity}
Exotic particles are necessary for anomaly cancellation in most $\uonep$ models. These are usually non-chiral under the Standard
Model gauge group, but chiral under $\uonep$. As discussed in Section \ref{sec:generic} our scenario typically involves exotic
chiral supermultiplets with supersymmetric masses in the 10-100 TeV range, such as the $D+D^c$ quark pairs or $E+E^c$ lepton pairs  in the model of Section \ref{sec:model}. Our focus is not on the specific model, but rather on the general $\zp$-mediation scenario,
so we will mainly comment on the more general case.

There are several possibilities for the lightest exotic scalar or fermion of a given type \footnote{For a recent general discussion, see \cite{Kang:2007ib}.}: (a) One is that it is absolutely stable. This possibility is severely constrained by cosmology and by direct searches for
heavy stable particles. However, it would be viable if the reheating temperature after inflation was sufficiently low~\cite{Giudice:2000ex}, i.e, $T_R < 20-200\GeV$ for an exotic mass in the 10-100 TeV range~\cite{Kudo:2001ie}. (b) The most
commonly studied case, especially for nonsupersymmetric models,
is that the exotic decays by mixing with ordinary quarks and leptons, allowing decays
such as $D\rightarrow (dZ, uW, dH)$~\cite{Barger:1985nq,Andre:2003wc}.
However, such mixings are often forbidden in supersymmetric $\uonep$ models,
at least at the renormalizable level, by  $\uonep$ and/or
$R$-parity conservation. For example, in the specific models in  Section \ref{sec:model}
there are no allowed renormalizable level operators that could lead to $D-d$ mixing.
However, $E-e$ mixing could be induced by a non-holomorphic soft operator
$L H_u^\ast E^c$ or a bilinear $Ee^c$, if present, for the $Q_Q=-Q_1/3$
model, or by $LLE^c$ or $Ec^c \nu^c$ operators for $Q_Q=(Q_2-Q_1)/6$.
The latter case would require spontaneous $R$-parity violation via the vevs of
a scalar $\nu$ or $\nu^c$.
(c) Another possibility is the existence of renormalizable-level couplings allowing
the direct decay of an exotic into ordinary particles,
such as the leptoquark (diquark) couplings $D u^c e^c$ ($D^c u^c d^c$)~ \cite{Kang:2007ib}.
One or the other could be present without inducing proton decay, and they would still allow
a stable LSP (the exotic scalar would be the normal particle). No such $D$ couplings are
allowed in the models in Section \ref{sec:model}, but analogous couplings for the $E$ or
$E^c$ (listed above in connection with mixing) could allow the rapid decays of $E$ and $E^c$
\footnote{The alternative models for a small neutrino mass do not allow
either possibility (b) or (c).}.
(d) Finally, exotics could decay by higher-dimensional operators, analogous to
(\ref{eqn:nuvariant}), which could induce highly suppressed mixing with the ordinary
particles or lead  directly to the decays. They would therefore be stable on collider time scales, leading
to exiting tracks or delayed decays in the detector~\cite{Kang:2007ib}.
Only dimension 5 operators would decay fast enough to satisfy constraints from
big bang nucleosynthesis~\cite{Kang:2007ib,Kawasaki:2004qu}. The only example in the models considered here
is $L H_d E^c S/M_P$, occurring in the $Q_Q=-Q_1/3$ case.

Thus, the lightest $D$ fermion or scalar would be stable in the specific models of
Sections \ref{sec:model} or \ref{section:alt}, which is unacceptable unless $T_R$ is very
low. However, such operators can always be allowed for both the $D$ and $E$-type exotics
by extending the particle content to include non-chiral exotics which obtain vevs.
We emphasise, however, that these models are only examples of a general scenario.

Finally, we comment briefly on $R$-parity, which is frequently
guaranteed by $\uonep$ invariance~\cite{Erler:2000wu}. In the present case, the operators
$S^n L H_u, S^n LL e^c, S^n u^c d^c d^c$, and $S^n QLd^c$, $n\ge 0$, are forbidden
for the specific  models considered in Section \ref{sec:model} by the $\uonep$ symmetry,
so there is an automatic $R$-parity in the Lagrangian, even after $\uonep$ breaking.
The alternative model in Section \ref{section:alt} with $Q_Q=-Q_1-2Q_2/3$ would allow the
$R$-parity violating operator $S u^cd^cd^c/M_P$. This operator would lead to LSP decay,
but with a lifetime much larger than (comparable to) the age of the universe for a wino (bino) LSP.

\subsection{Gauge Unification}
We comment briefly on gauge unification for the Standard Model couplings. The successful
unification in the MSSM is modified in the specific model considered here by the large Higgsino scale
and (especially) by the exotics. (This would be less of a problem in the $E_6$ motivated models,
which, however, lead to a bino LSP.)  The gauge unification could easily be restored by
additional non-chiral exotics, which could also lead to $\text{Tr} (QY)=0 $ at a high scale,
and possibly by a non-canonical normalization of the $U(1)_Y$ coupling~\cite{Barger:2006fm},
which occurs frequently in string constructions. As an example, approximate gauge unification at
around $3 \times 10^{15}\GeV$ would be achieved by the addition of four pairs
of $SU(2)$ doublets with $Y=0$ at around $2\times 10^{10}\GeV$. (These fractional charged
states could be confined at that scale.) We reemphasize that these
issues are very dependent on the specific model.


\begin{acknowledgments}
 We would like to thank Michael Dine, Aneesh Manohar, Nathan Seiberg, Jing Shao and Herman
 Verlinde for useful discussions.
  The work of L.W. and I.Y. is supported by the National Science
 Foundation under Grant No. 0243680 and the Department of Energy under
 grant \# DE-FG02-90ER40542. P.L is supported by the Friends of the
 IAS and by the NSF grant PHY-0503584. The work of G.P. was supported
 in part by the Department of Energy \# DE-FG02-90ER40542 and by the
 United States-Israel Bi-national Science Foundation grant \# 2002272.
Any opinions, findings, and conclusions or recommendations expressed
 in this material are those of the author(s) and do not necessarily
 reflect the views of the National Science Foundation.

\end{acknowledgments}

\appendix

\section{Charges and Anomaly Cancellation}
\label{app:anomalies}

In order to generate masses for all of the scalars, we assume that all of the visible sector
 chiral superfields, including the singlet,   are charged under $\uonep$. We also assume that there is
 only one singlet field $S$, and that there are no exotic $SU(2)$ doublets or additional Higgs pairs.
The charge assignments are constrained by the requirements of family universality, anomaly cancellation, and that
the superpotential terms in (\ref{eqn:superpotential}) and (\ref{eqn:exoticsuperpotential}) are allowed.
The
superpotential condition implies
\begin{eqnarray}
\left.\begin{matrix}
Q_2 + Q_Q + Q_{u^c} = 0 \\
Q_1 + Q_Q + Q_{d^c} = 0 \\
Q_2 + Q_L + Q_{\nu^c} = 0 \\
Q_1 + Q_L + Q_{e^c} = 0
\end{matrix}\right\} &~&\quad \text{Yukawa couplings}
\\\nonumber
\\
\left.\begin{matrix}
Q_S + Q_D+ Q_{D^c} = 0 \\
Q_S + Q_E + Q_{E^c} = 0 \\
\end{matrix}\right\} &~&\quad \text{Exotics couplings}
\\\nonumber
\\\label{eqn:SingletCoup}
Q_1 + Q_2 + Q_S = 0 \quad&~& \quad \text{Singlet coupling}
\end{eqnarray}

Based on the choice of exotics in the model in section
\ref{sec:model}, the anomaly cancellation conditions lead to the following constraints.

\textit{$SU_C(3)^2 \times \uonep$ anomaly cancellation:}
 \begin{equation}
 \label{eqn:su32u1}
n_D = 3.
\end{equation}

\textit{$SU_L(2)^2 \times \uonep$ anomaly cancellation:}
 \begin{equation}
 \label{eqn:su22u1}
Q_L = -3Q_Q - \frac{1}{3}(Q_1+Q_2).
\end{equation}

\textit{$\uonep$ gravitational anomaly cancellation:}
 \begin{eqnarray}
 \label{eqn:u1grav}
n_E= 2.
\end{eqnarray}

\textit{$U(1)_Y^2 \times \uonep$ anomaly cancellation:}
 \begin{eqnarray}
 \label{eqn:u1Y2u1}
 9 Y_D^2 + 2 Y_E^2 = 3,
\end{eqnarray}
where $Y_D = -Y_{D^c}$ and $Y_E = -Y_{E^c}$ are the hypercharges of
${D}$ and ${E}$.
  We will choose the hypercharges in analogy with the SM, $Y_D = -1/3$ and $Y_E = -1$.

\textit{$U(1)_Y \times U(1)^{\prime 2}$ anomaly cancellation:}
 \begin{eqnarray}
  Q_E=-3Q_Q -\frac{3}{2}Q_D + 2Q_1.
\end{eqnarray}

 \textit{$U(1)^{\prime 3}$ anomaly cancellation:}
 \begin{eqnarray}
 \label{eqn:u13}
&& 81Q_D^2-36Q_D(3Q_1+Q_2-3Q_Q)\\\nonumber
&+&4(7Q_1^2+8Q_1Q_2+Q_2^2-36Q_1Q_Q-27Q_Q^2)=0.
\end{eqnarray}
There are two possible choices for $Q_D$ as solutions to the
quadratic equation,
\begin{eqnarray}
 \label{cubicsoln}
Q_D &=& \frac{2}{9} (3 Q_1 + Q_2 -3 Q_Q)\\\nonumber &&\pm\sqrt{2(Q_1
+ 3Q_Q)(Q_1-Q_2 + 6Q_Q)}.
\end{eqnarray}
These correspond to two 2-parameter solutions in terms of $Q_2/Q_1$ and $Q_Q/Q_1$,
with the other charges obtained from the previous constraints.
Two simplified 1-parameter solutions are obtained by requiring
 the discriminant to vanish. We will mainly consider the case
\begin{equation}
\label{eqn:QQcharge} Q_Q = -\frac{1}{3} Q_1,
\end{equation}
and normalize $Q_1 = 1$,
so the other charges are all determined by $Q_2$, as listed in Table \ref{tab:charges}.

We note that
\begin{equation}
\label{eqn:crosstrace}
\text{Tr} (QY) =14Q_2-8Q_1+36Q_Q
\end{equation}
does not vanish in general, and not for the special 1-parameter solutions.
However, the vanishing can be restored by the addition of non-chiral states.
These do not affect the anomaly conditions and can also restore gauge
unification. 

\section{Renormalization Group Equations }
\label{app:RGEs} In calculating the various masses we distinguish
between two regions: $\mzp< \mu< \lams$ and $\mu< \mzp$. We use $t= \log
(\mu/\Lambda_S)$.
\subsection{$\mzp< \mu< \lams$}
For this region we use the RGEs given in
\cite{Yamada:1993ga,Martin:1993zk,Yamada:1994id}. To calculate the spectrum we need the one loop RGEs for the gauge and Yukawa couplings, $\zpino$, soft scalar masses, and the $A$ terms, as well as the two loop RGEs for the gaugino masses. 

Using $SU(5)$
normalization ($g_1^2=5g_Y^2/3$), the one loop $SU(3)_C \times SU(2)_L \times U(1)_Y$ gauge-couplings RGEs
are given by 
\begin{equation}
\frac{d g_a}{dt} = \frac{g_a^3}{16\pi^2}\, b_a,
\end{equation}
where $b_a=(51/5,1,0)$ for $a=1,2,3$. The $U(1)'$ gauge-coupling RGE is given by
\begin{equation}
\frac{d \gz}{dt} = \frac{\gz^3}{16\pi^2}\, {\rm Tr}\,Q_i^2.
\end{equation}

Keeping only the dominant terms proportional to $\mzp$, the two loop $SU(3)_C \times SU(2)_L \times U(1)_Y$ gaugino
RGEs are 
\begin{equation}
\frac{d \tilde{M}_a}{dt} = \frac{4 g_a^2  c_a}{(16\pi^2)^2}
\gz^2\mzp,
\end{equation}
with
\begin{eqnarray}
c_1 &=& \frac{6}{5}\sum_{\text{all scalars}} Q_i^2 Y_i^2 \nonumber\\
c_2 &=&  9 Q_Q^2 + 3 Q_L^2 +Q_1^2+Q_2^2 \\\nonumber
c_3 &=&  3(2 Q_Q^2 + Q_{u^c}^2 + Q_{d^c}^2) + n_D(Q_D^2 + Q_{D^c}^2).
\end{eqnarray}
The $U(1)'$ gaugino RGE is at one loop,
\begin{equation}
\frac{d \mzp}{dt} = \frac{\gz^2}{8\pi^2} \mzp{\rm Tr}\,Q_i^2.
\end{equation}
Within these approximations it is easy to  solve analytically the
gaugino RGEs. 

With the obvious definitions of the $A$ terms (see below (\ref{eqn:higgsMatrix})) their RGEs are 
\begin{eqnarray}
16\pi^2\frac{d {A}_D}{dt} &=&   4\gz^2 y_D (Q_D^2 + Q_{D^c}^2 + Q_S^2) \mzp \nonumber \\\nonumber
16\pi^2\frac{d{A}_E}{dt} &=&   4\gz^2 y_E (Q_E^2 + Q_{E^c}^2 + Q_S^2) \mzp  \\
16\pi^2\frac{d{A}_H}{dt} &=&   4\gz^2 \lambda (Q_{H_u}^2 + Q_{H_d}^2 + Q_S^2) \mzp, 
\end{eqnarray}
where we have neglected all the terms on the RHS that are not proportional to $\mzp$.

The RGEs for  the soft masses are 
\begin{eqnarray}
16\pi^2\frac{d m_S^2}{d t} &=&  - 8 \gz^2 Q_S^2 \mzp^2 
				+ 4\lambda^2(m_S^2 + m_{H_u}^2 + m_{H_d}^2) \nonumber \\\nonumber
                   &&   +6 n_D y_D^2(m_S^2 + m_D^2 + m_{D^c}^2) \\\nonumber&&+  2 n_E y_E^2(m_S^2 + m_E^2 + m_{E^c}^2) \\\nonumber
16\pi^2\frac{d {m}_D^2}{d t} &=&   -8 \gz^2 Q_D^2 \mzp^2 + 2y_D^2(m_S^2 + m_D^2 + m_{D^c}^2)  \\\nonumber
16\pi^2\frac{d {m}_{D^c}^2}{d t} &=&   -8 \gz^2 Q_{D^c}^2 \mzp^2 + 2y_D^2(m_S^2 + m_D^2 + m_{D^c}^2)  \\\nonumber
16\pi^2\frac{d {m}_E^2}{d t} &=&   -8 \gz^2 Q_E^2 \mzp^2 + 2y_E^2(m_S^2 + m_E^2 + m_{E^c}^2)  \\\nonumber
16\pi^2\frac{d {m}_{E^c}^2}{d t} &=&   -8 \gz^2 Q_{E^c}^2 \mzp^2 + 2y_{E^c}^2(m_S^2 + m_E^2 + m_{E^c}^2) \nonumber \\\nonumber
16\pi^2\frac{d {m}_{H_u}^2}{d t} &=&   -8 \gz^2 Q_2^2 \mzp^2 +2\lambda^2(m_S^2 + m_{H_u}^2 + m_{H_d}^2)\nonumber \\&&+ 6 Y_u^2(m_{H_u}^2 + m_{Q_3}^2 + m_{u_3^c}^2)   \nonumber\\\nonumber
16\pi^2\frac{d  {m}_{H_d}^2}{d t} &=&   -8 \gz^2 Q_1^2 \mzp^2 +2\lambda^2(m_S^2 + m_{H_u}^2 + m_{H_d}^2) \nonumber\\\nonumber&&+ 6 y_d^2(m_{H_d}^2 + m_{Q_3}^2 + m_{d_3^c}^2)   \\\nonumber
16\pi^2\frac{d {m}_{Q_3}^2}{d t} &=&   -8 \gz^2 Q_Q^2 \mzp^2 +2y_u^2(m_{H_u}^2 + m_{Q_3}^2 + m_{u_3^c}^2) \\\nonumber&&+  2y_d^2(m_{H_d}^2 + m_{Q_3}^2 + m_{d_3^c}^2)  \\\nonumber
16\pi^2\frac{d {m}_{u_3^c}^2}{d t} &=&   -8 \gz^2 Q_{u^c}^2 \mzp^2 +4y_u^2(m_{H_u}^2 + m_{Q_3}^2 + m_{u_3^c}^2)  \\\nonumber
16\pi^2\frac{d {m}_{d_3^c}^2}{d t} &=&   -8 \gz^2 Q_{d^c}^2 \mzp^2 +4y_d^2(m_{H_d}^2 + m_{Q_3}^2 + m_{d_3^c}^2),  \\
\end{eqnarray}
where we have ignored the (small) $A$ term contributions on the RHS \footnote{ In general,
there are also
   $U(1)_Y$ and $\uonep$ $D$-term contributions to
  the scalar RGEs, which are of the form Tr$(Y{m}_i^2)$ and Tr$(Q_i {m}_i^2)$. To
  order $\mathcal{O}(\gz^2)$ the contributions from the visible sector fields vanish in our scenario. Being the sum of scalar masses they vanish at the
  boundary $\mu = \Lambda_S$ like all the scalar masses. The RGE for
  this sum of masses is easily shown to be proportional to the sum
  itself by making use of the anomaly cancellation conditions on the
  charges. Being a homogeneous equation with vanishing boundary
  condition, the solution must vanish everywhere.
  Non-chiral hidden sector fields $\Psi$ and $\Psi^c$ could in principle
  yield non-vanishing $\uonep$ $D$-term contributions, but only if their soft mass-squares
  are unequal. Such effects would be of the same order as the $\zp$ contributions to the
  scalar masses.}.  

The one loop RGEs for the superpotential couplings are
\begin{eqnarray}
16\pi^2\frac{d \lambda}{d t}&=&\lambda\Big[4\lambda^2+3n_Dy_D^2+n_Ey_E^2+3y_u^2+3y_d^2\nonumber\\
&&-3g_2^2-\frac35 g_1^2-2\gz^2(Q_S^2+Q_1^2+Q_2^2)\Big]\nonumber\\
16\pi^2\frac{d y_D}{d t}&=&y_D\Big[2\lambda^2+(3n_D+2)y_D^2+n_Ey_E^2-\frac{16}{3}g_3^2\nonumber\\
&&-\frac65 g_1^2(Y_D^2+Y_{D^c}^2)-2\gz^2(Q_S^2+Q_D^2+Q_{D^c}^2)\Big]\nonumber\\
16\pi^2\frac{d y_E}{d t}&=&y_E\Big[2\lambda^2+3n_Dy_D^2+(n_E+2)y_E^2\nonumber\\
&&-\frac65 g_1^2(Y_E^2+Y_{E^c}^2)-2\gz^2(Q_S^2+Q_E^2+Q_{E^c}^2)\Big],\nonumber\\
\end{eqnarray}
with similar expressions for $y_u,y_d$, and $y_e$.
In practice, we ignored the relatively small running effects of $y_E$ and $y_e$.

To obtain the Higgs potential at the electroweak
scale one must run down below $\mzp$. In doing so, one encounters
several heavy thresholds. First, one must integrate out the  $\zp$-ino
and then the squarks, sleptons, Higgsinos, and exotics one by one. For simplicity and since
these masses are not greatly separated from $\mzp$ we will
ignore the running between these scales \footnote{This and other approximations we
  have made throughout, while
  small compared to the terms retained, are not negligible compared
  with the electroweak scale. However, the fine-tuning needed to
  obtain the electroweak scale is not restricted
  to a very small range of parameter space, so the approximations can be
  compensated by small changes in the values of the parameters such as
  $\lambda$ or exotic Yukawa couplings}.
\subsection{$\mu<\mzp$}
Below the mass scale of the scalars the Higgs mass and quartic's RGEs are those of
the Standard Model. (Unlike \cite{ArkaniHamed:2004fb}, there are no contributions 
from the Higgs-gaugino-Higgsino couplings in the low energy theory. 
Using the standard form of the Higgs potential, $m_H^2\phi^\dagger\phi+\lambda_H(\phi^\dagger\phi)^2/2$, we have \cite{Luo:2002ey},

\begin{eqnarray}
\label{eqn:RGE_H}
16\pi^2 \frac{d \lambda_H}{dt} &=& 12\left(\lambda_H^2+\lambda_Hy_t^2- y_t^4 \right)\nonumber\\
16\pi^2 \frac{d m_H^2}{dt} &=& 6m_H^2\left(\lambda_H + y_t^2\right),
\end{eqnarray}
where $y_t$ is the top Yukawa and we have neglected other smaller contributions. 

The RGE for $y_t$ is \cite{Kielanowski:2003jg},
\begin{eqnarray}
\label{eqn:RGE_Yt}
16\pi^2 \frac{d y_t}{dt} &=& \frac92 y_t^3-y_t\left(\frac{17}{20}g_1^2+\frac94g_2^2+8g_3^2\right)
\end{eqnarray}

We must also specify the matching conditions in passing from the high energy effective theory containing the  $\zp$-ino and scalars to the low energy theory with only SM fields and gauginos. The Higgs mass receives a quadratically divergent threshold correction from integrating out the squarks. The quartic coupling receives a contribution from the different $D$-terms as well as 
a contribution from an $F$-term,
\begin{eqnarray}
\label{eqn:quarticthreshold}
\lambda_H(\mu \approx {m}_{\varphi_i}) &=& \frac{1}{4}(g_2^2 + g_Y^2) + \gz^2 Q_2^2 +\frac12 \lambda^2 \sin^22\beta\nonumber\\
\label{eqn:massthreshold}
 m_H^2(\mu \approx {m}_{\varphi_i}) &=& min(\mathcal{M}_H^2) - \frac{3 y_t^2}{16\pi^2} m_{\varphi_i}^2.
\end{eqnarray}
Notice that the $F$-term contribution is small for large $\tan\beta$.

The gauge coupling RGEs in this region are 
\begin{equation}
\frac{d g_a}{dt} = \frac{g_a^3}{16\pi^2}\, b_a,
\end{equation}
where $b_a=(41/10,-11/6,-5)$. We do not run $\gz$.

The one loop  gauginos
RGEs in this region are
\begin{equation}
\frac{d \tilde{M}_a}{dt} = \frac{g_a^2}{16\pi^2}\tilde{M}_ac_a, 
\end{equation}
where  $c_a=(0,-12,-18)$.


\bibliographystyle{prsty}
\end{document}